# Tunable Sample-wide Electronic Kagome Lattice in Low-angle Twisted Bilayer Graphene

Qi Zheng[§], Chen-Yue Hao[§], Xiao-Feng Zhou, Ya-Xin Zhao, Jia-Qi He, Lin He[†]

Center for Advanced Quantum Studies, Department of Physics, Beijing Normal University, Beijing, 100875, People's Republic of China

[§]These authors contributed equally to this work.
[†]Correspondence and requests for materials should be addressed to Lin He (e-mail: helin@bnu.edu.cn).

**Overlaying two graphene layers with a small twist angle $\theta$ can create a moiré superlattice to realize exotic phenomena that are entirely absent in graphene monolayer. A representative example is the predicted formation of localized pseudo-Landau levels (PLLs) with Kagome lattice in tiny-angle twisted bilayer graphene (TBG) with $\theta < 0.3°$ when the graphene layers are subjected to different electrostatic potentials. However, this was shown only for the model of rigidly rotated TBG which is not realized in reality due to an interfacial structural reconstruction. It is believed that the interfacial structural reconstruction strongly inhibits the formation of the PLLs. Here, we systematically study electronic properties of the TBG with $0.075° \leq \theta < 1.2°$ and demonstrate, unexpectedly, that the PLLs are quite robust for all the studied TBG. The structural reconstruction suppresses the formation of the emergent Kagome lattice in the tiny-angle TBG. However, for the TBG around magic angle, the sample-wide electronic Kagome lattices with tunable lattice constants are directly imaged by using scanning tunneling microscope. Our observations open a new direction to explore exotic correlated phases in moiré systems.**

Atomic stacking configurations can have a profound influence on the electronic properties of graphene moiré systems [1-6]. In a graphene moiré structure, rotational alignment between two adjacent graphene sheets can introduce an SU(2) non-Abelian gauge field [7,8] and, at magic angle, give rise to flat bands [9,10], which now are an intriguing platform for exploring strongly correlated states and exotic quantum phases [3-6]. Besides the flat bands of magic-angle twisted bilayer graphene (MATBG), it was predicted that a perpendicular electric field can generate artificial gauge field and result in a new kind of flat band (pseudo-Landau levels) in tiny-angle TBG [11]. These pseudo-Landau levels (PLLs) are predicted to localize at the interfaces between AB and BA stacking regions with assuming a rigid rotation of two graphene layers and they form an emergent Kagome lattice in real space, possibly opening a new direction for exploring the physics of strong correlations. However, recent experiments demonstrated that the van der Waals (vdW) interlayer interaction causes atomic-scale reconstruction at the interfaces in small-angle TBG that minimizes the area of the energetically unfavorable AA stacking and enlarges AB and BA areas [12-17]. As a consequence, the true atomic stacking configuration in the tiny-angle TBG is quite distinct from a Kagome lattice [18,19]. Therefore, it is believed that the predicted PLLs and electronic Kagome lattice based on the assumption of a rigid rotation [11] will be strongly inhibited in the tiny-angle TBG [6].

In this Letter, we fabricate different TBG with $0.075° \leq \theta < 1.2°$ on insulating substrates [tungsten diselenide ($WSe_2$) and hexagonal boron nitride (h-BN)] and systematically study electronic properties of these TBG. Unexpectedly, our experiment demonstrates that the PLLs can be controllably switched on/off by using electric fields between two adjacent graphene sheets and they are quite robust for all the studied TBG. By using scanning tunneling microscope (STM), we directly measure spatial distribution of the PLLs and demonstrate that they are strongly localized in the interfaces, *i.e.*, domain walls (DWs), between the AB and BA regions in the reconstructed TBG. Although the reconstruction suppresses the formation of a structural Kagome lattice in tiny-angle TBG, localized PLLs are formed nevertheless. Only for larger twist angles the PLLs show the expected electronic Kagome lattice.

The Kagome lattice can host novel topological and strongly correlated states due to the destructive interference associated with frustrated geometry [20-26]. Figure 1(a) shows a schematic diagram of a two-dimensional (2D) Kagome lattice. Considering the nearest-neighbor hopping, quasiparticles are strongly confined in the frustrated Kagome lattice due to the completely destructive quantum phase interference (in the black dotted circle of Fig. 1(a)). Here, we point out that an electronic Kagome lattice naturally exists in the TBG. As illustrated in Fig. 1(b), there are three types of stacking configurations: AA stacking, AB/BA Bernal stacking, and DW transition regions, in the TBG. Obviously, the structure of the DW regions exhibit a Kagome lattice. Once there is a localized state in the DW regions, we can realize the electronic Kagome lattice embedded in the TBG.

To explore the electronic Kagome lattice in the TBG, different samples with controlled twist angles, ranging from 0.075° to 1.2°, are obtained by using transfer technology of graphene onto mechanical-exfoliated WSe$_2$ and h-BN sheets layer by layer [27-29] (see Fig. S1 and methods for details of the sample preparation [30]). The twist angle is estimated by $D = a/[2sin(\theta/2)]$, where $a = 0.246\ nm$ is the lattice constant of graphene and $D$ is the average wavelengths of moiré unit cells. Firstly, we study electronic properties of the tiny-angle TBG on the WSe$_2$ and h-BN. Figure 2(a) and 2(e) shows representative STM images of the tiny-angle TBG on the h-BN with $\theta < 0.2°$ (see Figs. S2-S5 for more experimental results of tiny-angle TBG on the h-BN [30]). The structural reconstruction in the TBG results in large triangular Bernal (AB and BA) stacking domains and small AA regions to minimize the total energy. A triangular network of the DWs separates the AB and BA domains and connects the different AA regions. To study effects of interlayer potential on the electronic properties of the tiny-angle TBG, we carry out scanning tunneling spectroscopy (STS) measurements to measure d$I$/d$V$ spectra by using two different tips. One is a tungsten (W) tip with negligible work function difference between the STM tip and the graphene [31,32], the other is an Au-coated W tip, as reported in Ref. [33], with a finite work function difference between the STM tip and the graphene. The Au-coated W tip induces an interlayer electric field in the graphene beneath the tip [33], which can lead to a gap in the AB/BA regions [34-38]. Figures 2(b) and 2(f) show representative STS spectra taken at the AA, AB/BA and DW regions of the TBG with the W tip and the Au-coated W tip, respectively. The d$I$/d$V$ spectra of the AA region recorded by both the tips show a pronounced peak, which is attributed to the AA modes, *i.e.*, the nearly-flat bands localized in the AA regions, of the tiny-angle TBG [11]. However, the spectra of the AB/BA and DW regions recorded by the two tips exhibit quite different features. For the W tip, gapless V-shaped spectra are observed in both the AB/BA and DW regions of different TBG [Fig. 2(b) and Figs. S2-S3]. Whereas, for the Au-coated W tip, the spectrum recorded in the AB/BA region exhibits a gap of about 50 meV (Fig. 2(f)). The gapped AB and BA regions have opposite valley Chern numbers, which generate topological helical edge states at the DWs [13,15-17,36]. By carrying out energy-fixed STS mapping at energies within the gap of the AB/BA regions, the one-dimensional (1D) topological edge states along the DWs are directly imaged in real space (see Fig. 2(g)). They are mainly located at the two edges of the DW, which is the characteristic feature of the topological states [13,15-17,36]. When there is no gap in the AB/BA regions, we cannot observe the 1D topological states (Fig. 2(c)). Similar results are observed for different tiny-angle TBG on the $WSe_2$ substrate, as shown in Figs. S6-S8 [30]. Our experiment indicates that the result is irrelevant of the twist angles in the TBG and is only determined by the tip-induced effective electric field.

Besides the 1D topological edge states, the STS measurement recorded by the Au-coated W tip reveals a new flat band localized in the DW (Fig. 2(f), at about 130 meV). Our energy-fixed d$I$/d$V$ map recorded at 130 meV indicates that the flat band is strongly localized in the center of the DWs, as show in Fig. 2(h) (see Figs. S5 for more experimental result of the ~0.12° TBG by using the Au-coated W tip). These features remind us the predicted PLL at the interfaces between the AB and BA stacking regions in the rigid-rotated tiny-angle TBG [11]. According to the bandgap ~ 50 meV in the AB/BA stacking regions, displacement field in the TBG is estimated as 0.5 Vnm$^{-1}$ and, therefore, the potential difference between the two adjacent graphene layers is estimated as about 0.165 V [34]. In the theory [11], when $\theta = 0.2°$ and $U = 0.165$ V, the energy of the PLL is calculated as about 0.135 eV, which agrees well with that observed in our experiment. Our experiment indicates that the structural reconstruction does not inhibit the formation of the PLL in the hole side. Such a result is reasonable because the structural reconstruction in the TBG is not expected to completely change the atomic structure at the interface between the AB and BA stacking regions [12-17]. The predicted PLL is strongly localized at the interface between the AB and BA stacking regions. Therefore, the localized gauge field at the interface generated by the interlayer electric field should still be observable. However, our experiment indicates that local strain induced by the structural reconstruction introduces strong electron-hole asymmetry [39,40] and suppresses the formation of the PLL in the electron side [Fig. 2(f) and Figs. S4-S7, no feature of the PLL is observed in the electron side in the measured range of energy]. Here, we should point out that it is usually difficult to observe higher energy PLL because that both the full width of the PLL and the signal of background increase with the energy in the STS measurements. In our experiment, the signal of the n = 1 PLL is quite robust and we cannot observe signal of the n = 2 PLL in all the studied TBG.

Our results, as shown in Fig. 2, demonstrate the formation of highly localized PLL, in addition to the helical network states, in the DWs of the tiny-angle TBG. However, because of the structural reconstruction, the structure of the tiny-angle TBG is completely distinct from that with assuming a rigid rotation of two adjacent graphene layers (Fig. 1(b)). Therefore, the highly localized PLL forms a triangular network connecting different AA regions (Fig. 2(h), Figs. S4 and S6), rather than a Kagome lattice as predicted in Ref. [11]. In the TBG, the structural reconstruction occurs for the twist angle below ~ 1.8° [14,17]. However, it will not completely change the Kagome structure of the DW regions in the TBG with the twist angle around the magic angle. To explore possible electronic Kagome lattice, we carry out measurements in the TBG with the twist angle around the magic angle by using the Au-coated W tip. Figure 3 summarizes the experimental results of three TBG samples with twist angles ~1.07°, ~0.98°, and ~0.88° (see Figs. S9-S17 for more experimental data [30]). It is quite surprising to find out that the localized PLL in the DWs is quite robust even in the TBG with the twist angle that is much larger than the tiny angle predicted in theory (Fig. 3(b)) [11]. By carrying out energy-fixed d$I$/d$V$ mappings at energies of the PLL, the emergent electronic Kagome lattices embedded in the TBG are directly imaged, as shown in Fig. 3(d). In our experiment, the PLL localized in the DWs is observed in all the studied TBG (see Figs. S9, S12 and S16 for more results). Still, no feature of the PLL is observed in the electron side in the measured range of energy, see Fig. S17 [30], indicating that the local strain induced by the structural reconstruction suppresses the formation of the PLL in the electron side. Besides the PLL, there are several wiggles in the spectra of Fig. 3, which are mainly localized either in the AB/BA regions or in the AA regions, indicating that they are arising from band edges of the AB/BA regions or arising from remote bands in the AA regions [41-47]. Figure 4(a) summarizes the measured energies of the PLL as a function of the rotation angles and the energy of the PLL increases with increasing the twist angle. For comparison, the theoretical evolution of the PLL (for four fixed perpendicular electric fields $0.05\ V \leq U \leq 0.2\ V$) as a function of twist angle in the TBG are also plotted. Even though the theoretical result is calculated according to the rigid-rotated TBG, the predicted increase of the energy of the PLL with twist angle agrees with that observed in our experiment. The difference is reasonably attributed to the structural reconstruction in the TBG, which is not taken into account in theory.

For the TBG with the twist angle around the magic angle, besides the PLL localized in the DWs, there are flat bands localized in the AA regions [41-47], as shown in Fig. 3(a) (see Figs. S9-S11 and S15 for more experimental results [30]). The bandwidth of the flat bands plays a vital role in the emergent strongly correlated states in the MATBG [3-6]. Figure 4(b) summarizes the measured full width at half maximum (FWHM) of the flat bands at the AA-stacked regions of TBG as a function of the rotation angles. Our experiment indicated that the bandwidth of the flat bands depends sensitively on the twist angle and reaches the minimum around the magic angle ~1.08°. The FWHM of the flat bands in the MATBG observed in this work is comparable to that reported in previous studies [41-47]. The calculated bandwidth of the low-energy flat bands using a continuum model of TBG [6,48] is also plotted for comparison. The experimental values of the bandwidth are much larger than the theoretical ones even with considering the broadening expected from experimental temperature (78 K), ~ 10 meV, and bias modulation, ~ 5 meV, suggesting that the broadening of flat bands observed in experiment is due to a loss of phase coherence of the electrons in the TBG.

Here, we should point out that the electronic Kagome lattice embedded in the TBG is quite different from the atomic Kagome lattice shown in Fig. 1(a) because that there are other electronic states, such as the AA modes, penetrating the electronic Kagome lattice. It will be of great interest to explore the interplay between the electronic Kagome lattice and other electronic states in the TBG. There are several advantages of the electronic Kagome lattices in the TBG: i) it is relatively easy to realize large-area, sample-wide Kagome lattice by fabricating graphene layer by layer; ii) the electronic Kagome lattices exist for a continuum set of angles around the magic angle; iii) the lattice constants of the Kagome lattices can be tuned by the twist angle. Therefore, the observed electronic Kagome lattices in the TBG may open a new direction to explore exotic correlated phases in moiré systems.

In summary, our experiments explicitly demonstrate that there is a robust and new type of flat band localized in the DW regions of small-angle TBG. In tiny-angle TBG, the highly localized flat band forms a triangular network connecting different AA regions because of the strong structural reconstruction. In the TBG for a wide range of twist angles around the magic angle, these localized states form electronic Kagome lattice with tunable lattice constants. The reported robust flat band and electronic Kagome lattice embedded in the TBG may open a new direction to explore exotic correlated phases in moiré systems.

**Acknowledgments:**

This work was supported by the National Key R and D Program of China (Grant Nos. 2021YFA1401900, 2021YFA1400100) and National Natural Science Foundation of China (Grant Nos. 12141401,11974050).

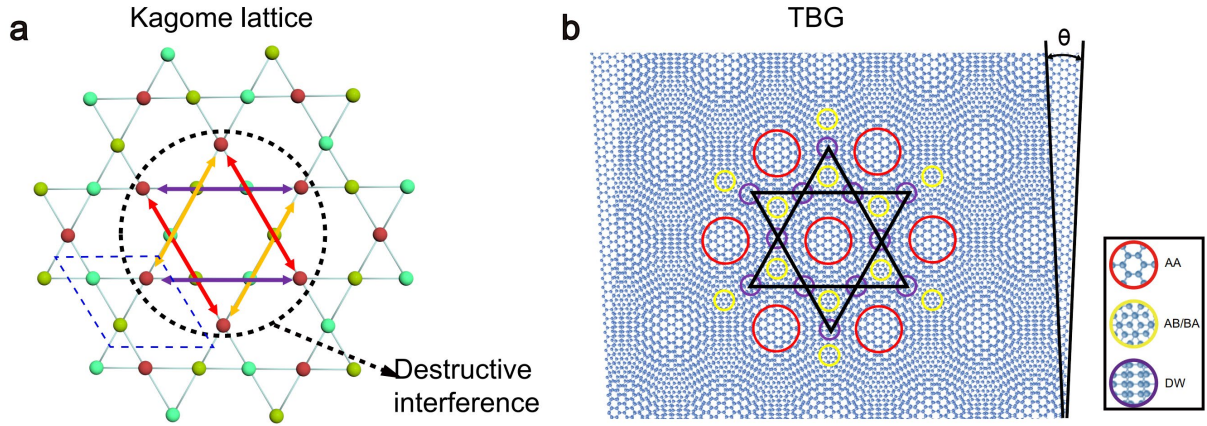

FIG. 1. Kagome lattice embedded in TBG. (a) Schematic diagram of destructive interference in the Kagome lattice. The blue dotted lines frame the region of an elementary cell, which consists of three lattice points (marked by dark green, green and brown). The red, yellow and purple arrows represent electron wave vectors in three different directions. The destructive interference of wave vectors at the brown lattice points results in the strongly localization of electrons in the black dotted circle. (b) Structural model of a TGB with rotation angle $\theta$. The AA, AB/BA, and DW stacking sites are marked by red, yellow, and purple circles, respectively. The DW sites exhibit the Kagome lattice.

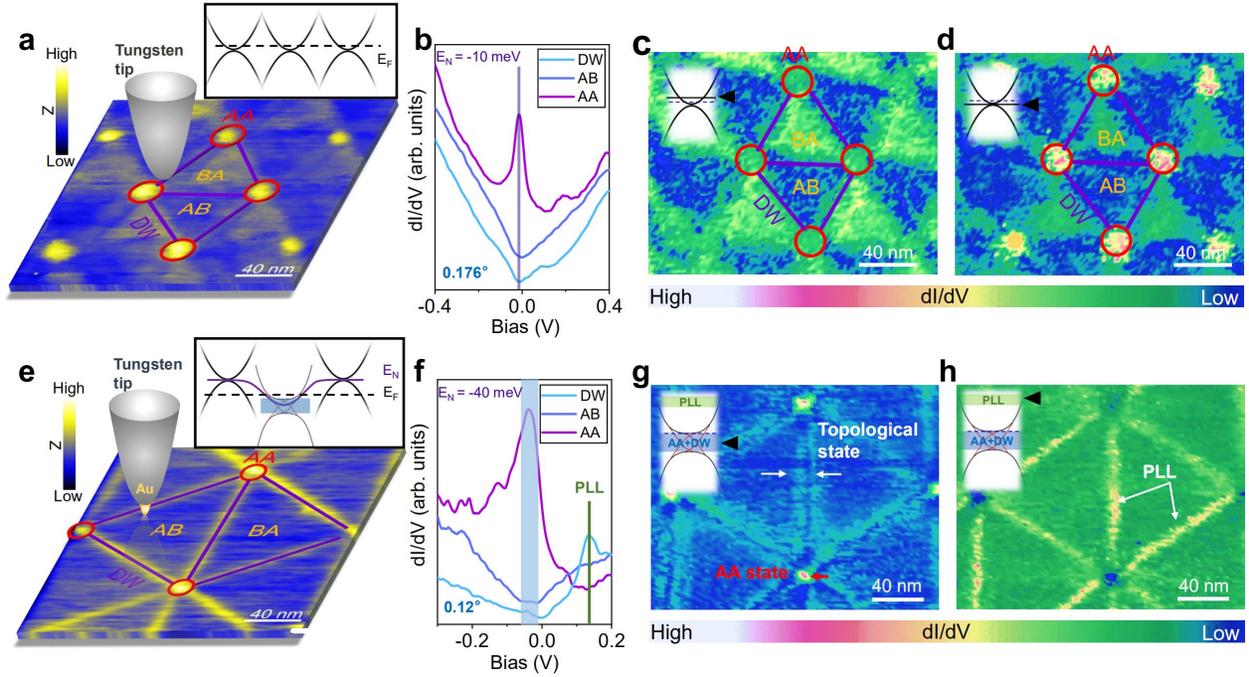

FIG. 2. STM/STS measurements of tiny-angle TBG on h-BN. (a) STM image of a 0.176° TBG measured by using the W tip. STM image ($V_{bias}$ = 500 mV, $I$ = 200 pA). Schematic bands along with the Fermi level (the dashed line) are plotted. (b) Representative STS spectra recorded in different stacking regions in (a). The CNP of the TBG is $E_N$ = -10 meV. (c) and (d) Energy-fixed d$I$/d$V$ maps at $E$ = 60 meV and -10 meV ($I$ = 150 pA). (e) STM image of a 0.12° TBG measured by using the Au-coated W tip. Schematic bands along with the Fermi level (the dashed line) and the charge neutrality point (CNP) of the TBG (the solid purple line) are plotted. The large work function difference between gold and graphene generates an interlayer electric field on the graphene below the tip. STM image ($V_{bias}$ = 500 mV, $I$ = 150 pA). (f) Representative STS spectra recorded in different stacking regions in (e). The blue rectangle denotes the bandgap in AB stacking sites. (g) and (h) Energy-fixed d$I$/d$V$ maps at $E$ = -40 meV and 130 meV.

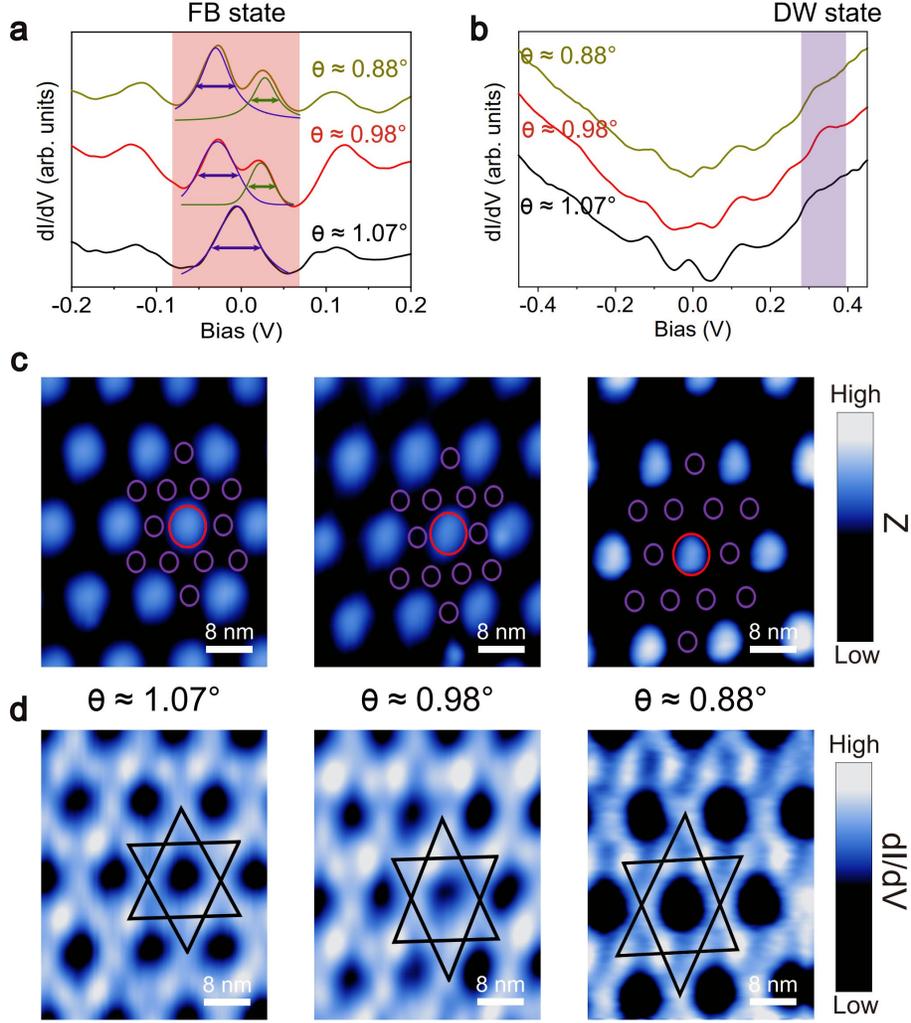

FIG. 3. Electronic Kagome lattice of the PLL states at the DWs in the low-angle TBG by using the Au-coated W tip. (a) d$I$/d$V$ spectra taken at AA-stacked regions of TBG with different rotation angles. At $\theta \sim 1.07°$, only a single peak is observed in the flat band (FB) energy region, whereas, there are two peaks at $\theta \sim 0.98°$ and $0.88°$. (b) d$I$/d$V$ spectra taken at the DW regions of the TBG with different rotation angles. DW states can be observed in all the studied TBG. (c) STM images of the low-angle TBG with different rotation angles ($V_{bias}$ = 500 mV, $I$ =100 pA). The bright regions in the red circles represent AA stacking sites and the purple circles represent the DW regions. (d) Energy-fixed d$I$/d$V$ maps at $V_{bias}$ = 380 mV in the TBG. Electronic Kagome lattices can be clearly imaged in all the three TBG.

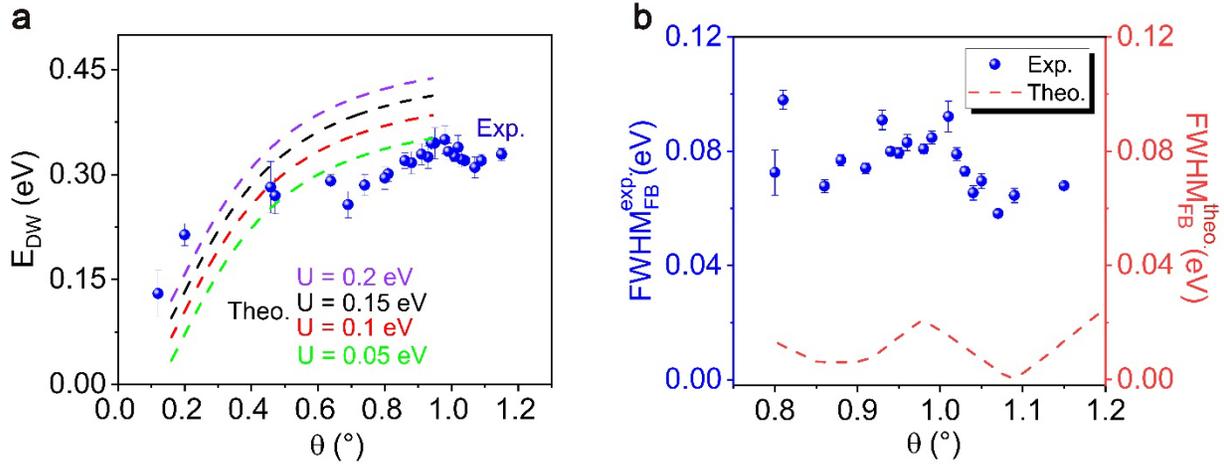

FIG. 4. Evolution of the DW states and the flat bands in AA regions as a function of twist angle. (a) The measured energies of the PLL (blue dots) at the DW regions as a function of the rotation angles. The dashed curves are the theoretical results extracted from Fig. 2b of ref. 11 with the fixed interlayer electric field $U = 0.05$eV (green), $U = 0.1$eV (red), $U = 0.15$eV (black), $U = 0.2$eV (purple). (b) The measured FWHM of the flat bands at the AA-stacked regions as a function of the rotation angles. The red dashed curve is the calculated bandwidth of the low-energy flat bands in the TBG. The FWHM of the DW states in the spectra was used to estimate the error bar in (a). The error bar in (b) is the fitted standard deviation.